\documentclass{article}

\usepackage[margin=1in]{geometry}
\usepackage[onehalfspacing]{setspace}
\usepackage{amsmath,mathtools}
\usepackage{amssymb}
\usepackage{pgfplots}
\usepackage{subcaption}
\usepackage{booktabs}
\usepackage[braket, qm]{qcircuit}
\usepackage{graphicx}
\usepackage{authblk}
\usepackage{xcolor}
\usepackage{booktabs}
\usepackage{subcaption}
\usepackage{comment}
\usepackage[linesnumbered,ruled,vlined]{algorithm2e}
\usepackage{qcircuit}
\usepackage{pgffor}
\usepackage{ifthen}
\usepackage{tcolorbox}

\SetCommentSty{mycommfont}

\SetKwInput{KwInput}{Input} %
\SetKwInput{KwOutput}{Output}  %
\SetKwProg{Fn}{Function}{}{}

\usepackage{siunitx} %

\usepackage{multicol}
\usepackage{centeredline}
\usepackage{nicematrix}
\usepackage{arydshln}

\usepackage{dsfont} 

\newcommand{\labs}[1]{\left| #1 \right|} %

\usepackage[english]{babel}

\usepackage{pdflscape}
\usepackage{fancyvrb}
\usepackage{calc}
\usepackage{rotating}
\usepackage{pgf,tikz}
\usepackage{mathrsfs}
\usetikzlibrary{arrows}
\usepackage{mathtools}

\usepackage{listings}
\usepackage[]{qcircuit}
\usepackage{braket}
\usepackage{mathtools}
\usepackage{varwidth}
\usepackage{caption}
\usepackage{subcaption}
\usepackage{graphicx}
\usepackage[export]{adjustbox}

\usepackage{float}

\makeatletter
\def\paragraph{\@startsection{paragraph}{4}%
	\z@\z@{-\fontdimen2\font}%
	{\normalfont\bfseries}}
\makeatother

\usepackage{graphicx}
\usepackage{pgffor}
\usepackage{tikz,tikz-cd}
\usetikzlibrary{backgrounds,fit,decorations.pathreplacing}  %
\usepackage{booktabs}
\usepackage{enumerate}

\usepackage{amssymb, amsmath, amsthm}

\usepackage{xypic}%
\usepackage{scalerel}%
\usepackage{ulem}

\usepackage{graphicx}  %
\usepackage{amsmath}%
\usepackage{amsfonts}  %
\usepackage{amsthm} %
\usepackage{xkeyval}%
\usepackage{amssymb}
\usepackage{enumerate} %
\usepackage{color}
\usepackage{breqn}  %
\usepackage{bm}  %
\usepackage{cite}

\allowdisplaybreaks[1]
\usepackage{nicefrac}
\usepackage{booktabs}

\usepackage{comment}

\usepackage{kpfonts}
\usepackage{stackengine}
\usepackage{calc}
\newlength\shlength
\newcommand\xshlongvec[2][0]{\setlength\shlength{#1pt}%
	\stackengine{-5.6pt}{$#2$}{\smash{$\kern\shlength%
			\stackengine{7.55pt}{$\mathchar"017E$}%
			{\rule{\widthof{$#2$}}{.57pt}\kern.4pt}{O}{r}{F}{F}{L}\kern-\shlength$}}%
	{O}{c}{F}{T}{S}}

\usepackage{nicematrix}

\usepackage{enumitem}

\usepackage{hyperref}
\hypersetup{
	colorlinks= true, %
	citecolor = red%
}

\graphicspath{{./figures/}}

\newcommand{\RN}[1]{%
	\textup{\uppercase\expandafter{\romannumeral#1}}%
}

\allowdisplaybreaks

\newtheorem{thm}{Theorem}[section]

\newtheorem{remark}[thm]{Remark}

\def\<{\langle}
\def\>{\rangle}

\numberwithin{equation}{section}

\usetikzlibrary{matrix,fit}

\usepackage{pgfplots}
\pgfplotsset{compat=1.17}

\makeatletter
\def\smallunderbrace#1{\mathop{\vtop{\m@th\ialign{##\crcr
				$\hfil\displaystyle{#1}\hfil$\crcr
				\noalign{\kern3\p@\nointerlineskip}%
				\tiny\upbracefill\crcr\noalign{\kern3\p@}}}}\limits}
\makeatother

\newcommand{\AWQPEs}{AWQPE}
\newcommand{\mList}{[m_1, m_2, \dots, m_B]} 
\newcommand{\nTotal}{n} 
\newcommand{\currentM}{m_i} 
\newcommand{\phiEst}{\phi_{\text{est}}}
\newcommand{\phiRaw}{\phi_{\text{raw}}}

\newcommand{\AmbFlags}{\mathcal{A}} 
\newcommand{\LastIdx}{\mathcal{S}_{\text{idx}}} 
\newcommand{\NumShots}{N_{\text{shots}}}
\newcommand{\NumTarg}{n_T}

\newcommand{\Threshold}{\epsilon}
\newcommand{\CurrentIter}{i}
\newcommand{\TotalBitsEstimated}{k}
\newcommand{\MostLikelyOutcome}{\mathbf{b}_{\text{ml}}}

\newcommand{\FlagAmb}{\text{flag}_{\text{amb}}}

\newcommand{\currentMStart}{\ensuremath{m_{\text{start}}}}
\newcommand{\AWQAE}{Adaptive Windowed Quantum Amplitude Estimation}
\newcommand{\AWQAEs}{AWQAE}

\makeatletter
\def\algocf@startfloat{
 \let\orig@float\relax
 \ifx\orig@float\undefined
  \let\orig@float\float
 \fi
 \orig@float
}
\makeatother

\newenvironment{proof-of-AWQPE}
  {\subsection*{Proof of \AWQPEs \, Ambiguity Resolution with Probabilistic Guarantees}}
  {}

\SetKwInOut{Input}{Input}
\SetKwInOut{Output}{Output}

\begin{document}
 \title{Modular Quantum Amplitude Estimation: A Scalable and Adaptive Framework}
	\author[1]{Alok Shukla \thanks{Corresponding author.}}
	\author[2]{Prakash Vedula}
	\affil[1]{School of Arts and Sciences, Ahmedabad University, India}
	\affil[1]{alok.shukla@ahduni.edu.in}
	\affil[2]{School of Aerospace and Mechanical Engineering, University of Oklahoma, USA}
	\affil[2]{pvedula@ou.edu}
	
\date{}

	\maketitle

\begin{abstract}
Quantum Amplitude Estimation (QAE) is a key primitive in quantum computing, but its standard implementation using Quantum Phase Estimation is resource-intensive, requiring a large number of coherent qubits in a single circuit block to achieve high precision. This presents a significant challenge for near-term Noisy Intermediate-Scale Quantum (NISQ) devices. To address this, we introduce the Adaptive Windowed Quantum Amplitude Estimation (AWQAE) framework, a modular, scalable and adaptive approach that decouples estimation precision from the number of physical qubits required in a single circuit. AWQAE operates by iteratively estimating the phase bits in small, fixed-size chunks, using a number of smaller, independent quantum circuits, which are amenable to parallel processing.
A key technical contribution of this work is introduction of a phase resolution circuit and an ancilla-guided mechanism that enables accurate chunk assignment and eigenphase reconstruction in the presence of multiple eigenstates.
This design is inherently NISQ-friendly, by lowering circuit depth and qubit count per block to reduce decoherence and noise effects. A key component of our approach is a robust classical post-processing algorithm that resolves measurement ambiguities that arise during the iterative process. This post-processing routine uses a least-significant-bit (LSB)-to-most-significant-bit (MSB) correction to reconstruct the full, high-precision phase estimate, ensuring accuracy. By combining a modular quantum-classical loop with an ambiguity-aware reconstruction method, AWQAE offers a powerful and flexible solution for performing high-precision QAE on resource-constrained quantum hardware. Our approach demonstrates enhanced scalability,  and adaptability, making it a promising candidate for practical applications of QAE in the NISQ era.
\end{abstract}

\section{Introduction}

Quantum amplitude estimation (QAE) \cite{nielsen2002quantum, brassard1998quantum, brassard2000quantum} is a fundamental quantum algorithm with broad applications, including financial risk analysis, Monte Carlo simulations and numerical integration \cite{woerner2019quantum, stamatopoulos2020option, montanaro2015quantum, shukla2025efficientpartialsum}. The objective of QAE is to estimate an unknown amplitude, which is encoded in the quantum state of the system. This is achieved by leveraging the properties of the Grover operator, $\mathcal{Q}$, which amplifies the probability of the ``good" state, $\ket{\psi_1}$. The eigenvalues of the Grover operator are directly related to the phase $\theta$, and estimating this phase allows us to determine the amplitude $a$.

Standard Quantum Amplitude Estimation (QAE), as originally proposed by Brassard et al.~\cite{brassard1998quantum, brassard2000quantum}, is a well-established algorithm based on the principles of Quantum Phase Estimation (QPE) \cite{nielsen2002quantum}. This approach estimates the eigenphase of the Grover operator to determine the desired amplitude, offering a quadratic speedup over classical methods. However, the resource requirements of standard QAE, including a large number of coherent qubits for the counting register and significant circuit depth, have motivated the development of alternative strategies better suited for Noisy Intermediate-Scale Quantum (NISQ) devices.

In response to these challenges, a variety of methods have emerged \cite{giurgica2022low}. Many approaches have moved away from the QPE formalism entirely \cite{suzuki2020amplitude, grinko2021iterative}. For instance, non-QPE-based methods often rely on statistical techniques like Maximum Likelihood Estimation (MLE) to infer the amplitude from measurement outcomes, as demonstrated in Ref.~\cite{tanaka2021amplitude}. Furthermore, some research methods have focused on adapting QAE to operate effectively in noisy environments \cite{herbert2024noise}.

We propose an enhanced approach to Quantum Amplitude Estimation (QAE), termed Adaptive Window Quantum Amplitude Estimation (AWQAE), which mitigates the resource-intensive requirements of standard QAE, namely, the need for a large number of coherent qubits in the counting register and considerable circuit depth. 
Our proposed method involves an adaptive, multi-qubit phase estimation algorithm that tackles the challenges of limited qubit resources and measurement ambiguities. The framework is characterized by a number of smaller, independent QPE-like circuits, each dedicated to estimating a specific block of phase bits. This modular and adaptive structure directly addresses several fundamental challenges associated with high-precision amplitude estimation. A principal strength of AWQAE lies in its scalability and resource efficiency. By decomposing the estimation process into smaller, independently executable blocks, AWQAE substantially reduces both the number of qubits and the circuit depth required at any given stage, leading to improved utilization of limited quantum resources and enhanced resilience to noise and decoherence.

The framework further demonstrates robustness to reconstruction ambiguities. It incorporates a statistical ambiguity check, complemented by a dedicated least-significant-bit to most-significant-bit correction routine in the post-processing phase. This combination ensures that the final estimate remains both accurate and reliable, even in the presence of inference uncertainties. Another notable feature of AWQAE is its adaptivity and configurability. The bit allocation strategy allows practitioners to finely adjust the algorithm's precision and resource demands to suit specific problem instances and hardware constraints, enabling a flexible trade-off between the number of estimation blocks and the complexity of each block.

The modular structure of AWQAE also facilitates a highly parallel and compartmentalized execution model, as each estimation block operates independently and can be executed concurrently across multiple quantum processors. This characteristic is especially well-suited to distributed quantum computing architectures. In addition, this modularity has the potential to result in significant tolerance to localized gate errors; errors within individual blocks do not propagate through the entire computational process. This potentiality localized error containment could enhance the robustness of the framework in noisy quantum environments.

The key contribution of this work is a novel algorithm (Algorithm \ref{alg:qae}) that performs the core task of amplitude estimation estimation in an adaptive, chunk-wise manner, and relies on a post-processing routine for ambiguity resolution (Algorithm \ref{alg:wiqpe_ambiguity_resolution}, \cite{shukla2025awqpe}) that applies a specialized correction to the raw binary phase estimate. This approach offer a powerful framework for performing high-precision QAE with a reduced number of physical qubits, making it a more viable solution for current and future quantum hardware.

Our proposed approach to \AWQAEs\ builds upon the Adaptive Windowed Quantum Phase Estimation (AWQPE) framework introduced in~\cite{shukla2025awqpe}. However, extending AWQPE to the quantum amplitude estimation (QAE) setting is non-trivial and constitutes a key technical contribution of this work. The original AWQPE algorithm assumes that the input state is an eigenstate of the unitary operator, whereas in the QAE setting, the input state is a superposition of eigenstates corresponding to two distinct eigenphases. This fundamental difference introduces significant challenges, as it is not immediately clear how to apply the chunked (or block-wise) processing of the original AWQPE when it is unknown which chunk belongs to which eigenphase.

In this work, we address this issue by introducing a phase resolution circuit and an ancilla qubit. The ancilla is flipped based on the most significant bit (MSB) of the phase resolution register, thereby encoding coarse phase information that guides the subsequent processing of chunks. This mechanism enables us to disambiguate the eigencomponents during reconstruction and accurately estimate the target eigenphase, effectively generalizing AWQPE to handle superpositions of eigenstates in the QAE framework.

\section{Standard Quantum Amplitude Estimation (QAE) Algorithm}

The standard Quantum Amplitude Estimation (QAE) algorithm uses Quantum Phase Estimation (QPE) to find the amplitude of a specific quantum state (Ref.~Algorithm \ref{alg:sqae}). The algorithm begins with a unitary operator $\mathcal{A}$ that prepares a quantum state $\ket{\psi}$ from the initial state $\ket{0}^{\otimes n}$, expressed as:
$$\mathcal{A}\ket{0}^{\otimes n} = \sqrt{1-p}\ket{\psi_0} + \sqrt{p}\ket{\psi_1}$$
Here, $\ket{\psi_1}$ represents the desired ``good" state, with success probability  $p = \sin^2(\theta/2)$, and $\ket{\psi_0}$ represents the orthogonal ``bad" states. 
The core of the QAE algorithm is the Grover operator, $\mathcal{Q}$, defined as $\mathcal{Q} = -\mathcal{A} S_0 \mathcal{A}^\dagger S_{\chi}$. Here,  $S_0 = I - 2\ket{0}^{\otimes n}\bra{0}^{\otimes n}$  and
 $S_{\chi} = I - 2\ket{\psi_1}\bra{\psi_1}$. This operator acts on the two-dimensional subspace spanned by $\ket{\psi_0}$ and $\ket{\psi_1}$. The eigenvalues of $\mathcal{Q}$ are $e^{\pm i\theta}$. The corresponding eigenstates of the Grover operator are:
$$\ket{\psi_\pm} = \frac{1}{\sqrt{2}} \left(\ket{\psi_0} \pm i\ket{\psi_1}\right)$$
These eigenstates have the eigenphases $\phi_\pm = \pm \frac{\theta}{2\pi}$. Note that the quantum state $\ket{\psi} $ can be expressed as a superposition of eigenstates of the Grover operator, as
\[
\ket{\psi} = \frac{1}{\sqrt{2}} \left( e^{-i \frac{\theta}{2}} \ket{\psi_+} + e^{i \frac{\theta}{2}} \ket{\psi_-} \right).
\]

The QAE algorithm uses QPE to estimate these eigenphases. A counting register of $m$ qubits is coupled to a target (or work) register containing the prepared state. The controlled powers of the Grover operator, $\mathcal{Q}^{2^j}$, are applied, and an Inverse Quantum Fourier Transform (IQFT) is performed on the counting register. The final measurement of the counting qubits yields an estimate of the phase $\phi$, from which 
 $\theta = \pm 2\pi \phi$ and  $\tilde{p} = \sin^2(\theta / 2)$
can be calculated. The algorithm provides an estimate with a precision of $\mathcal{O}(1/2^m)$, where  $m$ is the number of counting qubits.

\begin{algorithm}[H]
\caption{Quantum Amplitude Estimation (QAE) via Phase Estimation}
\label{alg:sqae}
\SetAlgoNlRelativeSize{-1}
\SetKwInOut{Input}{Input}
\SetKwInOut{Output}{Output}

\Input{
Unitary $\mathcal{A}$ such that $\mathcal{A} \ket{0}^{\otimes n} = \sqrt{p} \ket{\psi_1} + \sqrt{1 - p} \ket{\psi_0}$, where $\ket{\psi_1}$ is the ``good'' state.\\
Grover operator $\mathcal{Q} = -\mathcal{A} S_0 \mathcal{A}^\dagger S_{\chi}$, where:\\
\hspace{1em} $S_0 = I - 2\ket{0}^{\otimes n}\bra{0}^{\otimes n}$ reflects about the initial state,\\
\hspace{1em} $S_{\chi} = I - 2\ket{\psi_1}\bra{\psi_1}$ reflects about the ``good'' state $\ket{\psi_1}$.\\
Integer $m$: number of counting qubits (precision $\mathcal{O}(1/2^m)$).
}

\Output{
Estimate $\tilde{p}$ of the success probability $p = \sin^2(\theta / 2)$, with precision $\mathcal{O}(1/2^m)$.\\
(Optional) If the estimated amplitude is desired, return $ \sqrt{\tilde{p}}$.
}

\BlankLine

Initialize an $m$-qubit counting register in state $\ket{0}^{\otimes m}$ and an $n$-qubit target (or work) register in state $\ket{0}^{\otimes n}$\;

Apply $\mathcal{A}$ to the target register:
\[
\ket{0}^{\otimes n} \mapsto \sqrt{p} \ket{\psi_1} + \sqrt{1 - p} \ket{\psi_0}
\]

Apply Hadamard gates to the counting register\;

\For{$j = 0$ \KwTo $m-1$}{
 Apply controlled-$\mathcal{Q}^{2^j}$ to the target register, controlled by the $j$-th qubit of the counting register\;
}

Apply Inverse Quantum Fourier Transform (IQFT) on the counting register\;

Measure the counting register to obtain outcome  $y \in \{0, \dots, 2^m - 1\}$\;

Compute the phase $\phi = y / 2^m$\;

Compute $\theta = 2\pi \phi$ and set $\tilde{p} = \sin^2(\theta / 2)$\;

\textbf{return } $\tilde{p}$ (and optionally $\sqrt{\tilde{p}}$ as the estimated amplitude)\;

\end{algorithm}

\section{\AWQAE\  (\AWQAEs)}

We propose \AWQAE\, referred to as \AWQAEs,  a novel, scalable approach to quantum amplitude estimation by decoupling the precision of the estimate from the number of simultaneously coherent control qubits required in a single circuit block. The framework is characterized by a series of smaller, independent QPE-like circuits, each dedicated to estimating a specific block of phase bits. A robust classical post-processing routine is then applied to these partial and intermediate estimates to reconstruct the final, high-precision phase estimates and correct for potential ambiguities.

This modular architecture is particularly well-suited for Noisy Intermediate-Scale Quantum (NISQ) devices, as it reduces the required circuit depth and the total number of qubits in a single block. The overall process is detailed in Algorithm \ref{alg:qae}, with the ambiguity resolution post-processing specified in Algorithm \ref{alg:wiqpe_ambiguity_resolution}.

\subsection{The \AWQAEs\ Framework}

The input to Algorithm \ref{alg:qae} are as follows:  a unitary operator $\mathcal{A}$ that prepares the state $\mathcal{A} \ket{0}^{\otimes \NumTarg} = \sqrt{p} \ket{\psi_1} + \sqrt{1 - p} \ket{\psi_0}$, where $\ket{\psi_1}$ is the ``good'' state; the corresponding Grover operator $\mathcal{Q} = -\mathcal{A} S_0 \mathcal{A}^\dagger S_{\chi}$, with $S_0 = I - 2\ket{0}^{\otimes \NumTarg}\bra{0}^{\otimes \NumTarg}$ and $S_{\chi} = I - 2\ket{\psi_1}\bra{\psi_1}$; a bit allocation per block represented as an array $[m_1, m_2, \ldots, m_B]$ where each $m_i$ specifies the number of bits to be estimated in each of the $B$ iterative blocks, with the constraint that $m_i$ must be greater than $1$; and the number of measurement repetitions per iteration, $\NumShots$, which is critical for statistical confidence.

Upon completion, the algorithm yields the following outputs. First, it provides $\phiEst$, a final, corrected binary string that represents the full phase estimate after all ambiguities have been resolved. Second, it returns $\phiRaw$, the concatenated raw binary string of estimates from each block before any post-processing corrections have been applied. Third, it produces $\AmbFlags$, a list of boolean flags for each block that indicate whether the measurement outcome was considered ambiguous. Finally, the algorithm outputs $\LastIdx$, the index of a ``special chunk" if one was identified during post-processing, or \texttt{`None'} if no such chunk was found.

The initial phase of the algorithm sets up the necessary variables and registers for the iterative estimation process. The algorithm begins by initializing several key variables: $\phiRaw$, an empty string that will be used to concatenate the raw bit estimates from each block; $\nTotal$, the total number of bits to be estimated, calculated as the sum of all block sizes in the bit allocation list $[m_1, m_2, \ldots, m_B]$; $\TotalBitsEstimated$, a counter for the number of bits processed so far, initialized to 0; $\AmbFlags$, an empty list that will store a boolean flag for each block, indicating if the block's estimate was ambiguous; and $\CurrentIter$, an iteration counter starting from 0. An ambiguity threshold, $\Threshold$ (e.g., 0.9), is also defined to determine if two measurement outcomes are statistically too close to one another. The initial size for the phase resolution register, $\currentMStart$, is set to 2.

The main logic of the algorithm is encapsulated within a loop that continues as long as the total number of estimated bits, $\TotalBitsEstimated$, is less than the total number of bits to be estimated, $\nTotal$. Within each iteration of this loop, a new block of phase bits is estimated.

For each block, the algorithm first allocates a specific number of qubits for each of the logical registers: a phase resolution register with $\currentMStart$ qubits, a counting register with $\currentM$ qubits (the size of the current window), a target register with $\NumTarg$ working qubits, and a single ancilla qubit. Each of these qubits is initialized to the $\ket{0}$ state.

The quantum circuit for the block then begins with the preparation of the target register. The unitary operator $\mathcal{A}$ is applied to the target register, creating the initial superposition state $\ket{0}^{\otimes \NumTarg} \mapsto \sqrt{p} \ket{\psi_1} + \sqrt{1 - p} \ket{\psi_0}$. This is followed by the application of Hadamard gates to the phase resolution register. A sequence of controlled-$\mathcal{Q}^{2^j}$ operations is then applied to the target register, with the control qubits taken from the phase resolution register. This sequence prepares the state for the Inverse Quantum Fourier Transform (IQFT), which is subsequently applied to the phase resolution register. The final step of this preliminary part of the circuit involves a CNOT gate where the most significant bit (MSB) of the phase resolution register controls an ancilla qubit. This design ensures that the ancilla qubit's state encodes information about the phase. If the phase is in the range $[0, 1/2)$, the MSB is 0, leaving the ancilla in the $\ket{0}$ state; if the phase is in the range $[1/2, 1)$, the MSB is 1, flipping the ancilla to the $\ket{1}$ state. This encodes a high-level, single-bit estimate of the phase, which will be utilized later in the classical post-processing to guide subsequent corrections and ambiguity resolution.

Following the initial phase estimation using the phase resolution register, the algorithm proceeds to a more refined estimation using the counting register. Hadamard gates are applied to all qubits in the counting register, which prepares them for a subsequent phase estimation routine. This is followed by a series of controlled-$\mathcal{Q}$ operations, where the counting register's qubits control the application of the Grover operator $\mathcal{Q}$ to the target register. The exponent of $\mathcal{Q}$ is scaled by $2^{j+k}$ where $j$ is the index of the counting qubit and $k$ is an offset derived from the phase resolution register. This sequence of operations encodes the phase information into the counting register. The inverse Quantum Fourier Transform (IQFT) is applied to the counting register to map the encoded phase information into a measurable computational basis state.

After this, the ancilla qubit and the qubits of the counting register are measured in the computational basis. The entire quantum circuit for the block is executed $\NumShots$ times. During this process, measurement outcomes are only recorded if the ancilla qubit is measured to be in the  $\ket{b_a} = \ket{0}$ state; outcomes where the ancilla is  $\ket{b_a} =\ket{1}$  are discarded. This conditioning ensures that only measurements corresponding to the smaller of the two possible phase eigenvalues are considered, simplifying the subsequent classical processing.

The collected measurement results are then used to determine the most likely outcome, $t_1^*$, and the second most likely outcome, $t_2^*$. A classical post-processing step checks for ambiguity by comparing the counts of these two outcomes. If the ratio $C(t_2^*)/C(t_1^*)$ exceeds a predefined ambiguity threshold $\Threshold$, the block is flagged as ambiguous ($\FlagAmb \gets \text{True}$). In such a case, the final estimate for the block, $\MostLikelyOutcome$, is set to the modular minimum of $t_1^*$ and $t_2^*$ (referencing Equation \ref{eq:minabmod}). This specific choice addresses the cyclic nature of phase values at the boundaries. If the block is not ambiguous, $t_1^*$ is simply chosen as the estimate. The ambiguity flag and the selected estimate are then appended to their respective running lists, and the counters $\TotalBitsEstimated$ and $\CurrentIter$ are updated.

This iterative process continues until all blocks in the allocation list have been processed. Upon termination of the \texttt{While} loop, a final, comprehensive post-processing routine, \texttt{AWQPEAmbiguityResolution}, is called. This routine, which takes the concatenated raw estimates ($\phiRaw$), the bit allocation list, and the ambiguity flags as input, performs the necessary corrections to produce the final, unambiguous phase estimate $\phiEst$. Finally, the algorithm computes the phase angle $\theta = 2\pi\phiEst$ and from it, the final amplitude estimate $\tilde{p} = \sin^2(\theta / 2)$, which is then returned as the result of the algorithm.

\subsubsection{AWQPE Ambiguity Resolution}

The classical post-processing routine, \texttt{AWQPEAmbiguityResolution}, is a critical component of the \AWQAEs\ framework. Detailed in Algorithm \ref{alg:wiqpe_ambiguity_resolution}, this routine (originally presented in Ref.~\cite{shukla2025awqpe} and included here for completeness and ease of reference) takes the raw estimates ($\phiRaw$), the bit allocation list ($\mList$), and the ambiguity flags ($\AmbFlags$) as input to produce the final corrected binary phase estimate, $\phiEst$, and an index $\LastIdx$, which indicates a special chunk if present. Its primary function is to take the concatenated raw binary estimates ($\phiRaw$), the bit allocation list ($\mList$), and the ambiguity flags ($\AmbFlags$) as input, and produce a final, corrected binary phase estimate ($\phiEst$). The routine also identifies and returns the index ($\LastIdx$) of a special chunk if one is present.

The algorithm's first step is to logically partition the raw binary string $\phiRaw$ back into its original $m_j$-bit chunks, denoted as $[\phi^{(1)}, \phi^{(2)}, \ldots, \phi^{(B)}]$. It then performs a preliminary scan from the least significant bit (LSB) to the most significant bit (MSB) to identify a ``special chunk,'' defined as the rightmost non-zero chunk whose integer value is $2^{\currentM-1}$ (the binary string `10...0'). The index of this chunk is stored in $\LastIdx$ and is crucial for handling specific correction rules.

The core of the ambiguity resolution lies in a backward pass through these chunks, iterating from the second-to-last chunk to the first. In each step, a correction bit, $b_{\mathrm{corr}}$, is initially derived from the most significant bit of the next less significant chunk ($\phi^{(j+1)}$). The algorithm's robustness stems from its conditional application of this correction: the correction is bypassed (i.e., $b_{\mathrm{corr}}$ is set to 0) if the current chunk was flagged as ambiguous or if the next less significant chunk was identified as the special chunk. If neither of these conditions is met, the integer value of the current chunk is corrected by subtracting $b_{\mathrm{corr}}$ modulo $2^{m_j}$. After the correction is applied, the updated integer value is converted back to its $m_j$-bit binary string representation. This process continues until all chunks have been corrected. Finally, the individual corrected chunks are concatenated to form the final, unambiguous binary phase estimate, $\phiEst$.

The success of Algorithm~\ref{alg:qae} depends on accurately estimating the phase with high probability using our recently proposed Adaptive Windowed Quantum Phase Estimation (AWQPE) algorithm~\cite{shukla2025awqpe}. The correctness of AWQPE was established in Ref.~\cite{shukla2025awqpe}, which directly implies the correctness of Algorithm~\ref{alg:qae}.

\subsection{\AWQAEs\ Algorithm}

To facilitate modular comparisons in phase estimation, where values like $0$ and $n-1$ are adjacent in $\mathbb{Z}/n\mathbb{Z}$, we introduce a modified minimum function ~\cite{shukla2025awqpe}. For distinct $a, b \in \{0, 1, \ldots, n-1\}$, the modular minimum is defined as:
\begin{align} \label{eq:minabmod}
\min(a, b) \bmod n =
\begin{cases}
\min(a, b) & \text{if } a < b,\ a \notin \{0,n-1\},\ b \notin \{0,n-1\}, \\
n - 1 & \text{if } (a = 0 \text{ and } b = n - 1) \text{ or } (a = n - 1 \text{ and } b = 0).
\end{cases}
\end{align}
This definition captures the cyclic nature of modular arithmetic by treating $0$ and $n-1$ as neighbors, ensuring consistent behavior in edge cases.

\SetAlgoSkip{smallskip}
\SetAlCapSkip{1em}
\SetAlgoNlRelativeSize{-1}

\setlength{\algomargin}{1em} 

{\small
\IncMargin{-1.5em} 
\begin{algorithm}[H]
\caption{Quantum Amplitude Estimation (QAE) via Phase Estimation}
\label{alg:qae}
\SetAlgoNlRelativeSize{-1}
\SetKwInOut{Input}{Input}
\SetKwInOut{Output}{Output}

\Input{
Unitary $\mathcal{A}$ such that $\mathcal{A} \ket{0}^{\otimes \NumTarg} = \sqrt{p} \ket{\psi_1} + \sqrt{1 - p} \ket{\psi_0}$, where $\ket{\psi_1}$ is the ``good'' state.\\
Grover operator $\mathcal{Q} = -\mathcal{A} S_0 \mathcal{A}^\dagger S_{\chi}$, where
and $S_0 = I - 2\ket{0}^{\otimes \NumTarg}\bra{0}^{\otimes \NumTarg}$,  
\hspace{1em} $S_{\chi} = I - 2\ket{\psi_1}\bra{\psi_1}$.\\
Bit allocation per block $[m_1, m_2, \dots, m_B]$ with $m_i > 1$.
}

\Output{
Estimate $\tilde{p}$ of the success probability $p = \sin^2(\theta / 2)$, with precision $\mathcal{O}(1/2^n)$. Here $n = \sum_{i=1}^{B} m_i$. \\
(Optional) If the amplitude $\sqrt{p}$ is desired, return $\tilde{a} = \sqrt{\tilde{p}}$. 
}

\BlankLine
$\phiRaw \gets \texttt{""}$, $\nTotal \gets \sum_{j=1}^{B} m_j$, 
$\TotalBitsEstimated \gets 0$, 
$\AmbFlags \gets [\ ]$, 
$\CurrentIter \gets 0$,
set an ambiguity threshold $\Threshold$ (e.g., $0.9$), $\currentMStart \gets 2$, $b_a \gets 0 $; \\

\While{$\TotalBitsEstimated < \nTotal$}
{ 
 $\currentM \leftarrow $ the current window size for block $i$; \tcp{Current window size from $\mList$.}

Allocate $\currentMStart$ qubits to the phase resolution register, $\currentM$ qubits to a counting register,  $\NumTarg$ working qubits to a target (or work) register, and $1$ ancilla qubit. Initialize each of these qubits to $\ket{0}$ state;\\
 Apply $\mathcal{A}$ to the target register:
$\ket{0}^{\otimes \NumTarg} \mapsto \sqrt{p} \ket{\psi_1} + \sqrt{1 - p} \ket{\psi_0}$

Apply Hadamard gates to the phase resolution register\;
\For{$j = 0$ \KwTo $\currentMStart-1$}{Apply controlled-$\mathcal{Q}^{2^j}$ to the target register, controlled by the $j$-th qubit of the phase resolution register\;}
Apply Inverse Quantum Fourier Transform (IQFT) on the phase resolution register\;
Apply a CNOT gate on the ancilla qubit with the MSB qubit of the phase resolution register as the control;

Apply Hadamard gates to the counting register\;
\For{$j = 0$ \KwTo $\currentM-1$}{
  Apply controlled-$\mathcal{Q}^{2^{j+k}}$ to the target register, controlled by the $j$-th qubit of the counting register\;
}
Apply Inverse Quantum Fourier Transform (IQFT) on the counting register\;
Measure the ancilla qubit and counting qubits  in the computational basis;

Execute the circuit $\NumShots$ times and record the frequency of each outcome string of length $\currentM$ if the corresponding measurement for the ancilla qubit is $b_a $, and ignore the measurement outcome if the ancilla qubit is $(b_a + 1 \mod 2)$; 
 
  $C(t_1^{*}) \leftarrow $ the count of the most likely outcome, i.e., $C(t_1^*) \geq C(t)$ for $t \in \{0,\dots, 2^{m_i} -1\}$, where $C(t)$ is the count of observing the state $t$ on measurement 
  \tcp{Determine the most likely outcome. In case of tie, any choice is picked randomly.} 
  
 $C(t_2^{*}) \leftarrow $ the count of the most likely outcome, i.e., $C(t_2^*) \geq C(t)$ for $t \in \{0,\dots, 2^{\currentM} -1\} - \{t_1^*\}$; \\
 $\MostLikelyOutcome \leftarrow t_1^*$; \tcp{Select $t_1^*$ as the chunk estimate.}
  \tcc{Check for ambiguity.}
  $\FlagAmb \gets \text{False}$;\\
 \If{({${C(t_{2}^*)}/ {C(t_1^*)} > \Threshold$}) } 
 {
  $\FlagAmb \gets \text{True}$;\\
  \If{($\TotalBitsEstimated + \currentM < \nTotal$)}
  {
  $\MostLikelyOutcome \leftarrow  min(t_1^{*}, t_2^{*}) \mod 2^{\currentM}$; \tcp{Select $\min(t_1^*, t_2^*) \bmod 2^{\currentM}$ as the chunk estimate. Refer Eq.~\ref{eq:minabmod}.}
  }
 }
 Append $\FlagAmb$ to $\AmbFlags$;\\
 $\phiRaw \leftarrow \text{ concatenate } \phiRaw \circ \MostLikelyOutcome$; 
 \tcp{Here '$\circ$' operation represents concatenation of strings.}
 $\TotalBitsEstimated \gets \TotalBitsEstimated + \currentM$, 
 $\CurrentIter \gets \CurrentIter + 1$;
}

Apply a post-processing routine \texttt{AWQPEAmbiguityResolution}\((\phiRaw, \mList, \AmbFlags)\) to obtain 
$\phiEst$ and $\LastIdx$ \tcp*{Refer to Algorithm~\ref{alg:wiqpe_ambiguity_resolution}.}

Compute $\theta = 2\pi \phiEst$ and set $\tilde{p} = \sin^2(\theta / 2)$\;

\textbf{return } $\tilde{p}$ (and optionally $\sqrt{\tilde{p}}$ as the estimated amplitude)\;

\end{algorithm}
\DecMargin{-1.5em}  
}

\begin{algorithm}[H]
\small
\SetAlgoLined
\SetKwInOut{Input}{Input}
\SetKwInOut{Output}{Output}

\Input{
$\phiRaw$: Concatenated raw estimate from Algorithm~\ref{alg:qae}; \\
$ [m_1, m_2, \dots, m_B]$: Bit allocation per block;\\
$\AmbFlags$: List of ambiguity flags (MSB to LSB order).
}

\Output{
$\phiEst$: Final corrected binary phase estimate;\\
$\LastIdx$: Index of special chunk (if any), otherwise \texttt{`None'}.
}

\BlankLine
\tcp{Split concatenated string into $m_j$-bit chunks}
Partition $\phiRaw$ into substrings $[\phi^{(1)}, \phi^{(2)}, \dots, \phi^{(B)}]$ such that the size of the $j$-th block is $m_j$, i.e., $|\phi^{(j)}| = m_j$;\\
$\LastIdx \gets \texttt{`None'}$; 

\BlankLine
\tcp{Step 1: Identify rightmost non-zero chunk equal to $2^{m_j - 1}$ (i.e., '10...0')}
\For{$j \leftarrow B$ \KwTo $1$}{
 Let $x \gets$ integer value of $\phi^{(j)}$;\\
 \If{$x = 0$}{Continue;}
 \Else{
  \If{$x = 2^{m_j - 1}$}{
$\LastIdx \leftarrow j$; \\
}
 \textbf{break}
  }
 }

\BlankLine
\tcp{Step 2: Apply LSB-to-MSB corrections}
\For{$j \gets B - 1$ \KwTo $1$}{
 $b_{\mathrm{corr}} \gets$ most significant bit of $\phi^{(j+1)}$; \tcp{Default correction from MSB of next chunk}

 \If{($\AmbFlags[j] = \text{True}$) or ($\LastIdx = j+1$)}{
  $b_{\mathrm{corr}} \gets 0 $; \tcp{For ambiguous chunk, or if next chunk is special, no correction needed.}
 }

 Let $x \gets$ integer value of $\phi^{(j)}$;\\
 $x_{\mathrm{new}} \gets (x - b_{\mathrm{corr}}) \mod 2^{m_j}$; \tcp{Apply correction modulo $2^{m_j}$}
 $\phi^{(j)} \gets$ binary string of $x_{\mathrm{new}}$ with $m_j$ bits;
}

\BlankLine
\tcp{Step 3: Reconstruct corrected binary phase}
$\phiEst \gets$ concatenate $\phi^{(1)} \circ \phi^{(2)} \circ \cdots \circ \phi^{(B)}$;

\KwResult{$\phiEst$, $\LastIdx$}
\caption{\texttt{AWQPEAmbiguityResolution} - \AWQPEs \, Ambiguity Resolution (LSB-to-MSB Post-Processing)}
\label{alg:wiqpe_ambiguity_resolution}
\end{algorithm} 

Remarks 4.1 and 4.2 in Ref.~\cite{shukla2025awqpe} are also relevant to this work and are summarized below.

\begin{remark}[Special Chunk]
The special chunk ($\LastIdx$) handles the rare case where the fractional phase equals $0.5$, making the standard LSB-to-MSB correction ambiguous. Although unlikely, this can be addressed by slightly adjusting $m_i$ or perturbing the phase (see Remark~\ref{remark:perturb}). Even without correction, the measurement yields the correct bit with 50\% probability.
\end{remark}

\begin{remark}[Enhanced Confidence]\label{remark:perturb}
To resolve special chunk ambiguity, AWQPE can be rerun with a phase-perturbed unitary $U' = e^{i2\pi\Delta\phi} U$. Comparing the two estimates $\phi_{\text{est}}$ and $\phi'_{\text{est}}$ verifies correctness if $\phi'_{\text{est}} - \phi_{\text{est}} \approx \Delta\phi \pmod{2\pi}$.
\end{remark}

\begin{remark}
   It is important to note that the method presented here estimates two eigenphases (i.e., one less than $0.5$ and the other greater than $0.5$), using a single ancilla qubit. The most significant bit (MSB) of the phase resolution register controls a flip on the ancilla, effectively determining whether the phase lies below or above $0.5$. This decision then directs the subsequent circuit block to estimate the appropriate eigenphase.

In cases involving multiple eigenphases, and assuming prior knowledge of their separation, multiple ancilla qubits can be employed to distinguish between them. For example, consider four eigenphases $\phi_i$ for $i = 1$ to $4$, such that $\phi_i \in \left[\frac{i-1}{4}, \frac{i}{4}\right)$. In this case, the top two MSBs of the phase resolution register can be used to select the relevant eigenphase for further processing, allowing accurate estimation of the desired component.
\end{remark}

\begin{remark}
    In Algorithm \ref{alg:qae}, \currentMStart\ is initialized to $2$. While larger values could be used, they are unnecessary, as $2$ is sufficient for the phase resolution mechanism to function correctly. In contrast, setting \currentMStart\ to $1$ would not provide reliable performance.
\end{remark}

\begin{remark}
    In Algorithm \ref{alg:qae}, initially, $b_a$ could be set to either $0$ or $1$ and the same value is used throughout the algorithm. Both choices are generally valid; however, in rare edge cases where one setting yields no measurement outcomes (e.g., if $b_a = 0$ and $p = 1$), the alternative value should be used instead (i.e., $b_a = 1$ in this case) to ensure successful execution.
\end{remark}

\subsection{Computational Complexity}

The computational complexity of the \AWQAEs\ framework can be analyzed by considering its resource requirements for both quantum hardware and classical processing. The modular design fundamentally alters the resource trade-offs compared to standard Quantum Amplitude Estimation (QAE). For this analysis, let $n$ be the number of bits in the final phase estimate and $\NumTarg$ be the number of target qubits. We denote the gate complexity of the Grover operator, $\mathcal{Q}$, as $C_g(\mathcal{Q})$ and its circuit depth as $C_d(\mathcal{Q})$.

\subsubsection{Quantum Complexity}

The quantum complexity can primarily be quantified by the number of qubits, the circuit depth, and the total gate count.

\textbf{Qubit Resources:} Unlike conventional QAE, which requires a number of counting qubits equal to the desired precision $n$, the \AWQAEs\ algorithm partitions the problem into a series of smaller blocks. For any given block $i$, the total number of qubits required is limited to the sum of the qubits for the phase resolution register, the counting register, the target register, and a single ancilla qubit, which is $\currentMStart + m_i + \NumTarg + 1$. Consequently, the maximum number of qubits needed at any single time is determined by the largest block size, $\max(m_i)$, which can be significantly smaller than the total number of bits required for the final precision $n$. This design makes the algorithm highly resource-efficient for devices with a limited number of coherent qubits.

\textbf{Circuit Depth and Gate Count:} The total number of quantum gates is dominated by the controlled applications of the Grover operator, $\mathcal{Q}$. The \AWQAEs\ approach distributes the total computational effort of a standard QAE circuit across multiple shallow circuits. While the total number of applications of $\mathcal{Q}$ across all blocks is $O(2^n)$.
However, one key advantage of our proposed lies in the circuit depth. The depth for $j$-th block is given by  $O(2^{(-1 + \sum_{r=1}^{j} m_r )} \cdot C_d(\mathcal{Q}))$. This is  an improvement over standard QAE (which has a quantum circuit depth of $O(2^n C_d(\mathcal{Q}))$ ). We note that in the standard QAE, the quantum circuit is highly susceptible to decoherence. By distributing the computation into multiple shallower circuits, \AWQAEs\ could potentially offers a more resilient architecture for execution on noisy hardware.

\subsubsection{Classical Complexity}

The classical complexity of \AWQAEs\ is primarily associated with the post-processing routine, \texttt{AWQPEAmbiguityResolution}. This routine involves logical operations on the measurement outcomes, including partitioning the raw binary string $\phiRaw$ and a subsequent LSB-to-MSB correction loop. These operations have a computational cost that scales linearly with the total number of bits estimated, $\nTotal$. Given that these classical operations are highly efficient on modern processors, the post-processing overhead is negligible compared to the quantum circuit execution and does not represent a limiting factor for the algorithm's overall performance.

\subsubsection{Advantages of the \AWQAEs\ Framework}
The \AWQAEs\ framework offers several distinct advantages over conventional Quantum Amplitude Estimation (QAE) techniques, establishing it as a compelling approach for implementation on both current and future quantum hardware. Its modular and adaptive structure directly addresses several fundamental challenges associated with high precision amplitude estimation.

A principal strength of \AWQAEs\ lies in its scalability and resource efficiency. Standard QAE protocols require a large number of coherent counting qubits to achieve high precision, often resulting in quantum circuits that exceed the capabilities of present day Noisy Intermediate Scale Quantum (NISQ) devices. By decomposing the estimation process into smaller, independently executable blocks, \AWQAEs\ substantially reduces both the number of qubits and the circuit depth required at any given stage. This leads to improved utilization of limited quantum resources and enhances the overall resilience of the algorithm to noise and decoherence.

The framework further demonstrates robustness to reconstruction ambiguities. It incorporates a statistical ambiguity check, complemented by a dedicated least significant bit to most significant bit correction routine in the post processing phase. This combination ensures that the final estimate remains both accurate and reliable, even in the presence of inference uncertainties.

Another notable feature of \AWQAEs\ is its adaptivity and configurability. The bit allocation strategy, encoded by the sequence $[m_1, m_2, \ldots, m_B]$, allows practitioners to finely adjust the algorithm’s precision and resource demands to suit specific problem instances and hardware constraints. This enables a flexible tradeoff between the number of estimation blocks and the complexity of each block. For instance, fewer but larger blocks may be advantageous for systems equipped with stable and high fidelity qubits, whereas a greater number of smaller blocks may be preferable under more restrictive  conditions.

The modular structure of \AWQAEs\ also facilitates a highly parallel and compartmentalized execution model. Each estimation block operates independently and can be executed concurrently across multiple quantum processors.

Taken together, the theoretical strengths of the \AWQAEs\ framework, including modular structure and resource efficient execution, underscore its suitability for near term quantum platforms. The subsequent section presents a detailed evaluation based on numerical simulations, providing empirical validation of its practical performance.

\section{Results}
\label{sec:results}

The performance of the proposed Adaptive Windowed Quantum Amplitude Estimation (\AWQAEs) algorithm was evaluated by comparing its amplitude estimates to those obtained via the standard full Quantum Amplitude Estimation (QAE) algorithm across a set of representative and random test cases. In each trial, the true amplitude was analytically defined as $\sqrt{p} = \labs{\sin(\pi \phi)}$,
where \( p \) is the probability of measuring the “good” state and \( \phi \in [0, 1) \) is the encoded eigenphase. Both \AWQAEs\ and full QAE were used to estimate \( \phi \) ($\phiEst$ in Algorithm \ref{alg:qae}), from which the estimated amplitude  $\sqrt{\tilde{p}}$ was obtained.

To ensure a fair comparison, we excluded all trials for which the \texttt{special\_chunk\_flag} was set to \texttt{True}, as such cases may involve ambiguous bit recovery. 

\subsection{Amplitude Estimation Accuracy}

Table~\ref{tab:results} presents a selection of test cases. For each instance, the estimated amplitude from \AWQAEs\ (using $[m_1, m_2, m_3] = [3, 3, 4]$))  is shown alongside the corresponding value from full QAE (using $10$ counting qubits). The percentage error is computed relative to the full QAE baseline. As shown, all estimates from \AWQAEs\ match those from full QAE to within numerical precision, yielding a relative error of zero in every case.

\begin{table}[H]
\centering
\caption{Comparison of amplitude estimates between \AWQAEs\ (using $[m_1, m_2, m_3] = [3, 3, 4]$)) and full QAE (using $10$ counting qubits).}
\label{tab:results}
\begin{tabular}{ccccc}
\toprule
\textbf{Trial} & \textbf{True Amplitude} & \textbf{Est. Amplitude (\AWQAEs)} & \textbf{Est. Amplitude (Full QAE)} & \textbf{Error (\%)} \\
\midrule
1 & 0.9233 & 0.9239 & 0.9239 & 0.00 \\
2 & 0.1542 & 0.1528 & 0.1528 & 0.00 \\
3 & 0.7460 & 0.7451 & 0.7451 & 0.00 \\
4 & 0.9524 & 0.9524 & 0.9524 & 0.00 \\
5 & 0.4708 & 0.4714 & 0.4714 & 0.00 \\
6 & 0.8168 & 0.8176 & 0.8176 & 0.00 \\
7 & 0.1815 & 0.1800 & 0.1800 & 0.00 \\
8 & 0.4081 & 0.4080 & 0.4080 & 0.00 \\
9 & 0.7939 & 0.7940 & 0.7940 & 0.00 \\
10 & 0.4243 & 0.4248 & 0.4248 & 0.00 \\
\bottomrule
\end{tabular}
\end{table}

\subsection{Key Observations}

The results in Table~\ref{tab:results} demonstrate that the \AWQAEs\ algorithm achieves amplitude estimates that are indistinguishable from those of the full QAE method in all the cases. This confirms that the modular, windowed structure of \AWQAEs\ does not compromise estimation fidelity.

Moreover, by attaining comparable accuracy with a significantly lower resource footprint, achieved through independent blockwise estimation using fewer qubits and reduced circuit depth, \AWQAEs\ presents a promising alternative for deployment on Noisy Intermediate-Scale Quantum (NISQ) hardware, where circuit depth and qubit coherence are limiting factors.

\begin{figure}[h!]
    \centering
    \begin{subfigure}[b]{0.9\textwidth} 
        \centering
        \includegraphics[width=0.9\textwidth]{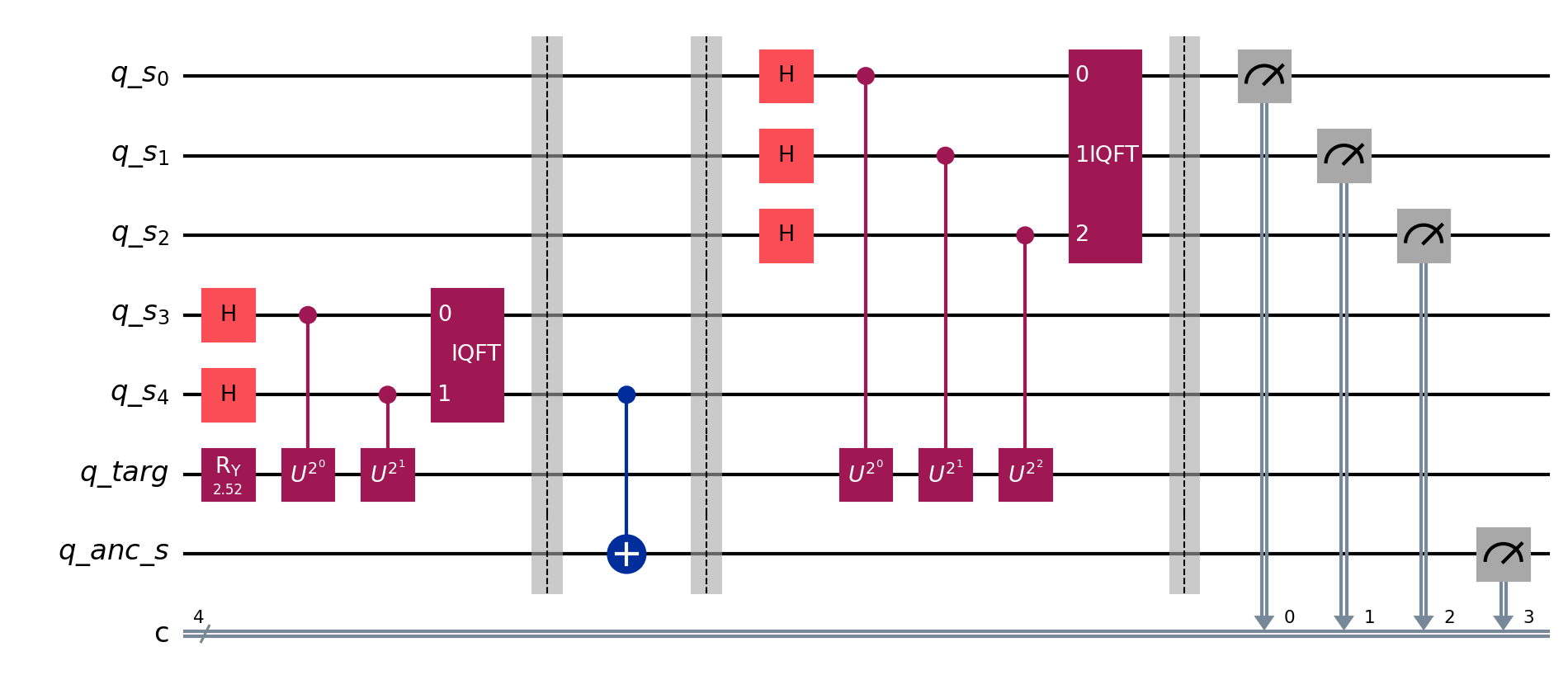} 
        \caption{Circuit for the first iteration block, estimating 3 most significant bits (MSBs).}
        \label{fig:circuit_iter1}
    \end{subfigure}
    \vspace{0.1cm} 

    \begin{subfigure}[b]{0.9\textwidth}
        \centering
        \includegraphics[width=0.9\textwidth]{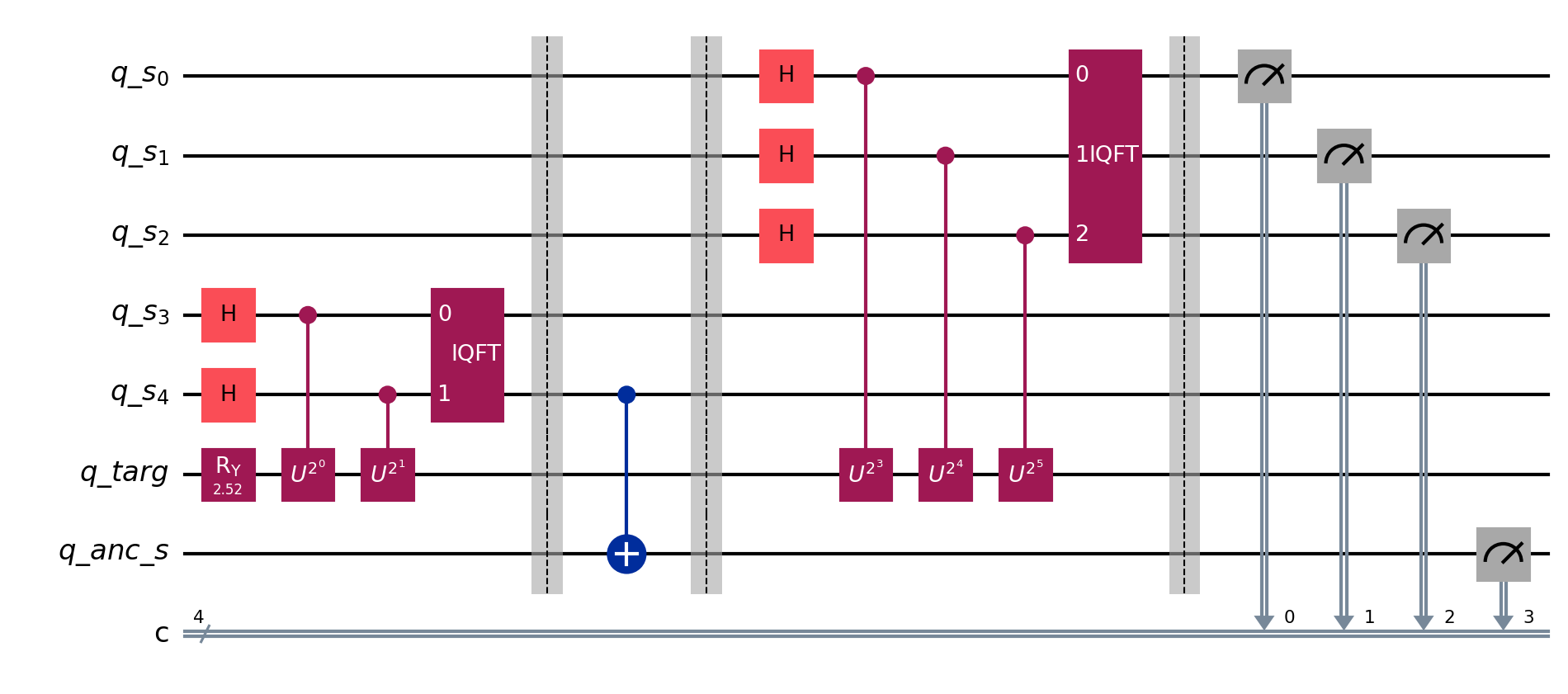} 
        \caption{Circuit for the second iteration block, estimating the next 3 bits.}
        \label{fig:circuit_iter2}
    \end{subfigure}
    \vspace{0.1cm} 

    \begin{subfigure}[b]{0.9\textwidth}
        \centering
        \includegraphics[width=0.9\textwidth]{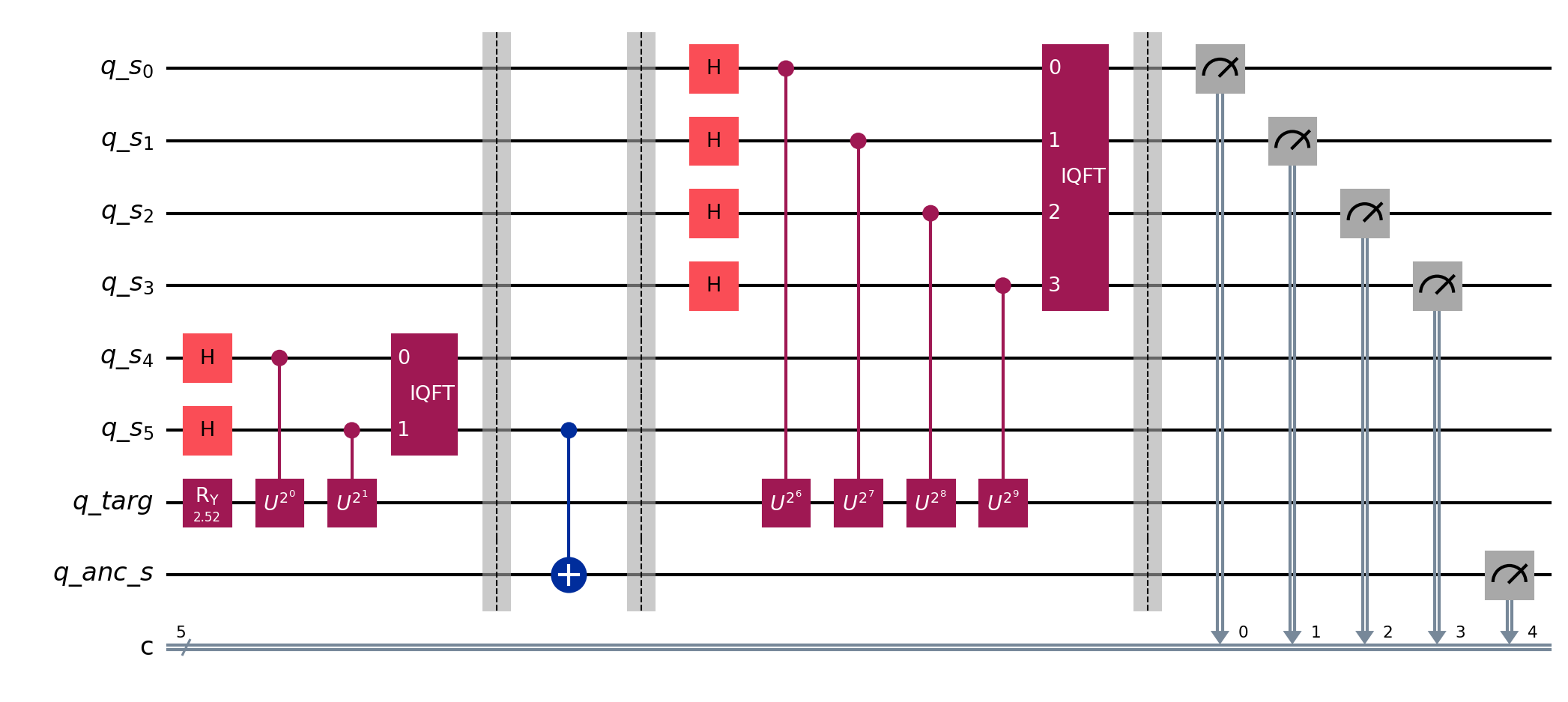} 
        \caption{Circuit for the third iteration block, estimating the final 4 bits.}
        \label{fig:circuit_iter3}
    \end{subfigure}

    \caption{Quantum circuit diagrams for the non-uniform chunk size test case (with a true amplitude of $a_{true} = \sqrt{p} = 0.9523504170755709$ using $[m_1, m_2, m_3] = [3, 3, 4]$). Each circuit corresponds to an iterative block of the \AWQAE \, algorithm (Algorithm \ref{alg:qae}).}
    \label{fig:all_circuits}
\end{figure}

\begin{figure}[h!]
    \centering
        \centering
        \includegraphics[width=0.9\textwidth]{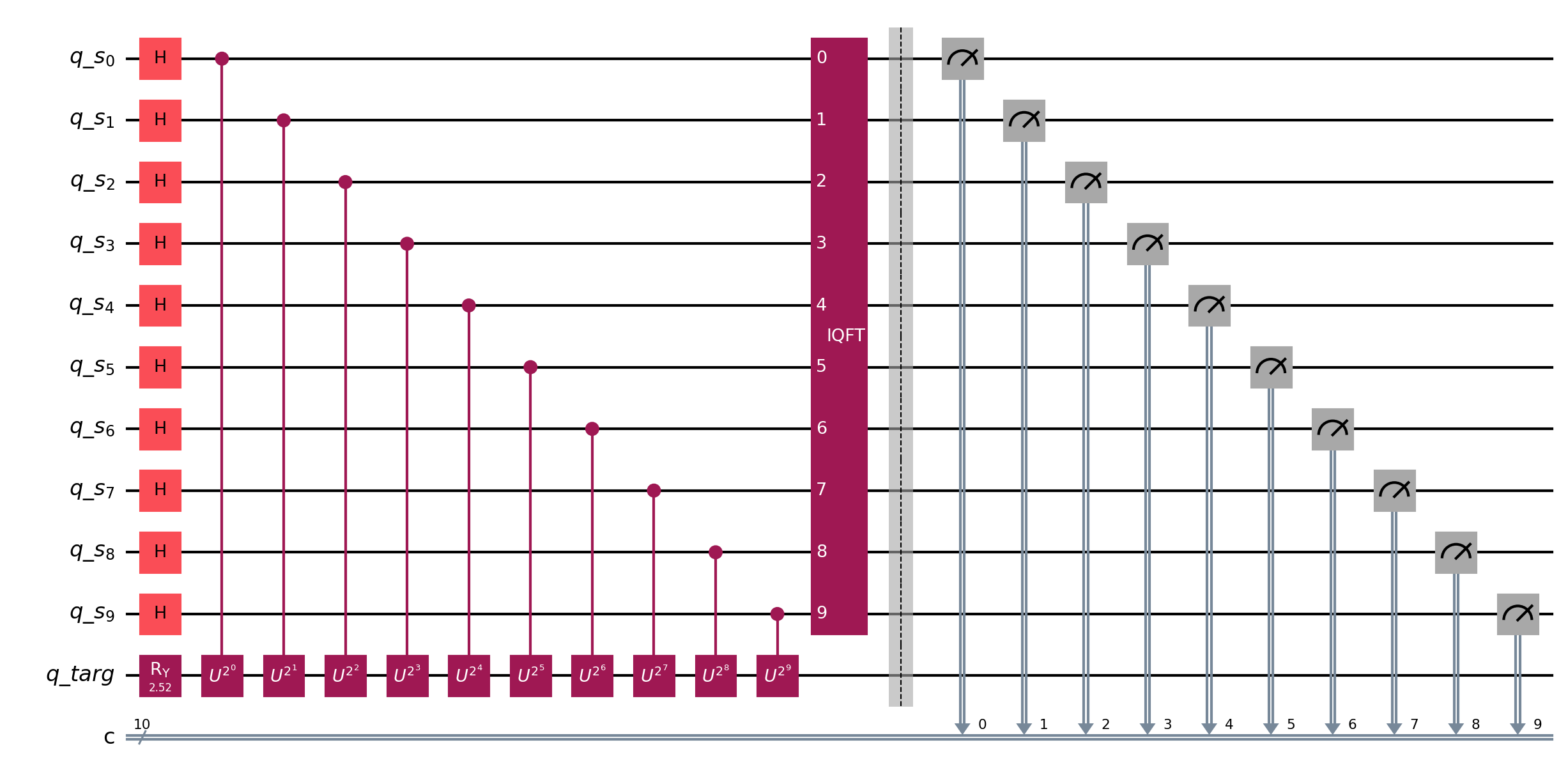} 
        \caption{Quantum circuit for the standard QAE for the  test case (with a true amplitude of $a_{true} = \sqrt{p} = 0.9523504170755709$ with $n = 10$ counting qubits).}
        \label{fig:circuit_full}
    \end{figure}

\paragraph{Circuit Structure and Modular Estimation Layout.}
Figure~\ref{fig:all_circuits} presents the quantum circuit diagrams corresponding to a representative test case of the \AWQAEs\ algorithm, executed with a true amplitude of \( a_{\text{true}} = \sqrt{p} = 0.9523504170755709 \). The total precision of ten bits is distributed across three independent estimation blocks, using a non-uniform bit allocation given by \( [m_1, m_2, m_3] = [3, 3, 4] \). Each block enables an intermediate estimate of a distinct contiguous segment of the phase bits.

The first block, shown in Figure~\ref{fig:circuit_iter1}, estimates the three most significant bits using three counting qubits. The Grover operator \( \mathcal{Q} \) is applied with powers \( 2^0, 2^1, 2^2 \), followed by an Inverse Quantum Fourier Transform (IQFT) and measurement. The second block, shown in Figure~\ref{fig:circuit_iter2}, continues the estimation with powers \( 2^3, 2^4, 2^5 \). The third block, shown in Figure~\ref{fig:circuit_iter3}, completes the sequence by estimating the four least significant bits using powers \( 2^6 \) through \( 2^9 \).

This modular decomposition enables each estimation block to operate independently, allowing for parallel execution and targeted error mitigation. Because the circuit depth per block remains shallow, the overall resource complexity scales more favorably with desired precision. The measurement of the ancilla qubit enables post-selection on successful amplitude amplification outcomes.
Estimates of the amplitude obtained from our \AWQAEs\ and the standard QAE were found to be equal ($\sqrt{\hat{p}} = 0.9523750127197659$).

For comparison, Figure~\ref{fig:circuit_full} shows the circuit for the full QAE algorithm with ten counting qubits. This circuit applies \( \mathcal{Q} \) in powers from \( 2^0 \) to \( 2^9 \), followed by a global inverse QFT across all qubits. While the full QAE circuit achieves high-fidelity amplitude estimation, it incurs significantly greater circuit depth and resource demand. In contrast, \AWQAEs\ achieves equivalent resolution with substantially reduced overhead, making it better suited to near-term quantum hardware.

\subsection{Relevance to Grover’s Algorithm and Quantum Counting}

The AWQAE algorithm (similar to the standard QAE algorithm) is also relevant  to quantum counting and Grover’s search \cite{grover1996fast, brassard1998quantum, shukla2025efficientsearch}, both of which rely on phase estimation of the Grover operator $\mathcal{Q}$. In the context of quantum counting, the goal is to estimate the number of marked items $M$ in an unstructured database of size $N$. This is achieved by estimating the amplitude $a = \sqrt{M/N} = |\sin(\pi \phi)|$, where $\phi$ is the eigenphase associated with the Grover operator acting on the “good” subspace. The modular estimation procedure introduced in AWQAE can be directly applied in this setting, enabling resource-efficient estimation of $\phi$ and, by extension, the count $M$, even when $M$ is initially unknown.

This connection is also highly relevant for Grover’s algorithm, which classically assumes knowledge of the number of marked items to optimize the number of Grover iterations. In many practical applications, however, this quantity is not known a priori, and over- or under-rotation can significantly reduce the success probability. By incorporating AWQAE as a preparatory step, one can efficiently estimate the amplitude and calibrate the number of Grover iterations accordingly. These observations underscore the practical utility of AWQAE as a general-purpose amplitude estimation primitive, especially in realistic, NISQ-era implementations of search and decision algorithms.

\section{Conclusion}

In this work, we introduced the Adaptive Windowed Quantum Amplitude Estimation (\AWQAEs) framework, a novel and highly scalable approach designed to overcome the resource limitations of standard Quantum Amplitude Estimation (QAE) on Noisy Intermediate-Scale Quantum (NISQ) devices. By moving away from the monolithic circuit structure of traditional QPE-based QAE, \AWQAEs\ effectively decouples the desired estimation precision from the number of simultaneously coherent qubits required in a single execution block. This is achieved through an iterative, multi-qubit phase estimation scheme that processes the phase bits in small, independent manageable chunks or blocks. If required, these blocks can also be processed in parallel.

The core of our framework lies in two key algorithms: a modular quantum loop for phase bit estimation (Ref.~Algorithm \ref{alg:qae}) and a powerful classical post-processing routine for ambiguity resolution (Ref.~Algorithm \ref{alg:wiqpe_ambiguity_resolution}). A key technical innovation of this work is the non-trivial extension of our Adaptive Windowed Quantum Phase Estimation approach (presented in Ref.~\cite{shukla2025awqpe})  to handle superpositions of eigenstates in QAE, achieved by introducing a phase resolution circuit and an ancilla qubit that guides chunk assignment for accurate eigenphase reconstruction. The iterative quantum procedure is inherently NISQ-friendly, as it uses smaller circuits with reduced depth, thereby mitigating the detrimental effects of decoherence and noise. The accompanying post-processing algorithm is essential for the framework's reliability, as it correctly identifies and resolves ambiguities that naturally arise in the measurement outcomes, ensuring that a high-fidelity, complete phase estimate can be reconstructed from partial measurements. Our results demonstrate that \AWQAEs\ offers potentially compelling advantages in terms of enhanced scalability,  and adaptability to various hardware constraints.

\bibliographystyle{unsrt}

\end{document}